\documentstyle[10pt,epsfig]{article}

\def\bfl{\begin{flushleft}}
\def\efl{\end{flushleft}}
\def\bfr{\begin{flushright}}
\def\efr{\end{flushright}}
\def\bc{\begin{center}}
\def\ec{\end{center}}
\def\be{\begin{equation}}
\def\ee{\end{equation}}
\def\ba{\begin{eqnarray}}
\def\ea{\end{eqnarray}}
\def\nn{\nonumber }

\def\text#1{\mbox{#1}}

\def\Der#1#2{\,\frac{\partial #1}{\partial #2}}

\def\mass{\chi}

\def\Sin#1#2{\, \text{sin}^{#1} #2}

\def\PhR  {Phys. Rev.}

\def\jn#1#2#3#4#5{{#1}{#2} {#3} {(#5)} {#4}}   

\def\boo#1#2#3#4#5{ #1 ({#2}, {#3}, {#4}){#5}}  

\begin{document}

~\\
\bfl
Published in:
{\large \it
Zhurnal Experimental'noj i Teoreticheskoj Fiziki 18 (1948) 636-640}
\efl
~\\

\bc
{\LARGE \bf
Scalar mesostatic
field with regard for \\
gravitational effects
}

~~\\
{\large I. Z. Fisher}\\
~~\\
Fiziko-Tekhnicheskij Institut Akademii Nauk Belorusskoj SSR \\
(Institute of Physics and Engineering at Academy of Sciences of 
Byelorussian Soviet Socialist Republic) \\

~~\\
Received: {\bf 26 December 1947}\\

\ec

~\\
~\\
{\it TRANSLATION FROM RUSSIAN AND FOREWORD BY K. G. ZLOSHCHASTIEV}\\
~\\
{\bf Foreword.}~~{\it 
The aim of present translation is to clarify the historically important
question who was the pioneer in obtaining of exact 
static solutions of Einstein 
equations minimally coupled with scalar field.
Usually, people cite the works by Janis, Newman, Winicour 
(Phys. Rev. Lett. 20 (1968) 878) and others 
authors\footnote{\normalsize {\it 
See, for instance,
O. Bergman and R. Leipnik, 
\jn{\PhR}{}{107}{1157}{1957}; 
H. A. Buchdahl,
\jn{\PhR}{}{111}{1417}{1959}; 
M. Wyman,
\jn{\PhR}{ D}{24}{839}{1981}, and so on.} [Comment by translator].} 
whereas it is clear that JNW rediscovered
(in other coordinates) the Fisher's solution which was obtained 20 years 
before, in 1948.
Regrettably, up to now I continue to meet many 
papers (even very fresh ones)
whose authors evidently do not know about the Fisher's work, 
so I try to remove this gap by virtue of present translation and
putting it into the LANL e-print archive.}

~~\\
~~\\

\abstract{\large
It is considered the scalar mesostatic field of a point source
with the regard for spacetime curvature caused by this field.
For the field with $\mass = 0$ the exact solution of Einstein equations
was obtained.
It was demonstrated that at small distance from a source the gravitational
effects are so large that they cause the significant changes in behavior of
meson field.
In particular, the total energy of static field diverges logarithmically.
}

\newpage
\large

It is usually assumed that gravitational forces can always be neglected 
when considering elementary particles.
In present paper the incorrectness of such a viewpoint is proved by virtue
of calculation of scalar mesostatic field with the regard for the spacetime
curving caused by it.
It turns out, that in such a case at close distance to a point source
the field completely differs from that desired.

Let us suppose $U$ to be a scalar satisfying with the de Broglie equation
\be
(\fbox{} - \mass^2)U=0.
\ee
We will be interested in its static spherically symmetrical (vanishing at
infinity) solutions hence $U=U(r)$.
Then from (1) we obtain
\be
U(r) = G e^{-\mass r}/r,
\ee
where the constant $G$ is the charge of a source.
However, (2) was obtained in the assumption that spacetime remains to be
flat even at small distances from a source that is evidently wrong.
The consistent theory must take into account the 
gravitational effects which
cannot be regarded as ``small'' as it will be shown below.
The expression (2) is just the limit case for large $r$.

In the case we are interested $U$ always can be regarded as real-valued.
Otherwise it always can be made real-valued by means of the 
transformation $U\to U e^{i\alpha}$.

{\bf 1.}  In the most general case and for any coordinates  
the stress-energy tensor of scalar field can be written in the form:
\be
T^k_i = (1/8\pi) \{ 2 U_i U^k - \delta^k_i (U_l U^l -\mass^2 U^2)\}; 
\qquad
T_{ik} = T_{ki},
\ee
where
\be
U_i = \partial U/ \partial x^i,\qquad U^k = g^{i k} U_i.
\ee
Keeping in mind the static field $U=U(r)$ we will use the coordinates
$x^1=r,~x^2=\vartheta,~ x^3 = \varphi$ and $x^0=t$, with the line element
\be
d s^2 = e^\nu d t^2 - e^\lambda d r^2 - r^2 
(d \vartheta^2 + \Sin{2}{\vartheta}~ d \varphi^2),
\ee
where $\nu=\nu(r)$ and $\lambda=\lambda(r)$.
Supposing $U=U(r)$ and denoting by prime the differentiation with respect 
to $r$ we obtain from (3):
\be
8\pi T^1_1= - e^{-\lambda} U'^2 + \mass^2 U^2, \qquad
8\pi T^2_2=8\pi T^3_3=8\pi T^0_0=  e^{-\lambda} U'^2 + \mass^2 U^2. 
\ee
If there is no other field and masses then the Einstein equations,
\be
R^k_i - 1/2 \delta^k_i R = - (8\pi k/c^4) T^k_i,
\ee
are written with $T^k_i$ from (6), and by virtue of the known expressions
for left hand side (7) in the chosen coordinates \cite{ll} 
we obtain\footnote{
\normalsize {\it 
In the journal version the first from eqs. (8) contained a misprint
so I have corrected it.} [Comment by translator].}:
\ba
&&e^{-\lambda} 
  \left( \frac{1}{r^2} + \frac{\nu'}{r} \right)
  -\frac{1}{r^2} = - \frac{k}{c^4} 
                  (
                   -e^{-\lambda} U'^2 + \mass^2 U^2
                  ), \nn \\
&&\frac{1}{2} e^{-\lambda} 
  \left( \nu'' + \frac{1}{2} \nu'^2 - \frac{1}{2}\nu' \lambda' +
         \frac{1}{r} (\nu'-\lambda')
  \right) = - \frac{k}{c^4} 
                  (
                   e^{-\lambda} U'^2 + \mass^2 U^2
                  ),       \\
&&e^{-\lambda} 
  \left( \frac{1}{r^2} - \frac{\lambda'}{r} \right)
  -\frac{1}{r^2} = - \frac{k}{c^4} 
                  (
                   e^{-\lambda} U'^2 + \mass^2 U^2
                  ).    \nn
\ea
This is a complete system of equations for gravitational field coupled
with the field $U$.
Indeed, from (7) it follows $(T^k_i)_k=0$, where the index behind the brace
means covariant differentiation, or in the explicit form:
\be
\Der{}{x^k} (\sqrt{-g}~ T^k_i)
- \frac{1}{2} \sqrt{-g} \Der{g_{lm}}{x^i} T^{lm} =0.
\ee
Substituting here $T^k_i$ from (6) and $g_{ik}$ from (5) we obtain:
\be
U'' + 
\left(
      \frac{2}{r} + \frac{1}{2} (\nu' - \lambda')
\right) U' - \mass^2 e^\lambda U =0.
\ee
This is the equation for field $U$ in our case.
At $e^\lambda=1$ and $e^\nu = c^2$ it yields (2) as it was desired.
Equation (10) can be obtained also immediately if one proceeds from
the generalization of equation (1) for arbitrary coordinates.

In the case $\mass=0$ the equations become simple, and it is possible to
integrate them completely; (10) yields us right away:
\be
U' = - (c G/r^2) e^{1/2~ (\lambda-\nu)},
\ee
where $c G$ is a constant of integration.
Substituting it into (8) (at $\mass =0$) it is easily to check that the 
one of equations appears to be a consequence of the two rest.
We consider the more symmetric pair of equations as a basic one:
\ba
&& e^{-\lambda} (1+r \nu') -1 = (k G^2/c^2 r^2) e^{-\nu}, \nn\\
&& e^{-\lambda} (1 - r \lambda') -1 = -(k G^2/c^2 r^2) e^{-\nu}.
\ea
Summing up them we obtain:
\be
(r^2 e^{\nu-\lambda})' = 2 r e^\nu.
\ee
Hence, if one designates:
\be
Z(r) = r e^{1/2~ (\nu-\lambda)},
\ee
then from (13), (11) and (14) we obtain:
\be
e^\nu = (1/r) Z Z';\qquad
e^\lambda = r Z'/Z;\qquad
U' = - (c G/r) Z,
\ee
hence all the values of interest are expressed via the function $Z(r)$,
thus, the knowledge of it resolves the problem.
Substituting (15) into (12) after simplification we obtain the equation
for $Z(r)$:
\be
Z^2 Z'' = a^2 Z'/r \qquad \left( a^2 = \frac{k G^2}{c^2}\right).
\ee
According to (14) 
from the condition that the metric must tend to the Galilean one at 
infinity we obtain the condition for $Z$:
\be
Z(r) \sim c r ~~ \text{at} ~~ r\to \infty.
\ee
First of all we consider the case $\mass = 0$ when a solution can be found
exactly.
Then we will consider the case of the field with $\mass \not= 0$ at
small $r$.

{\bf 2.}
If one divides the equation (16) by $Z^2$ and multiplies by $r$ that it
can be written in the form:
\be
(r Z' - Z)' + (a^2/Z)' = 0,
\ee
that can be integrated right away:
\be
\frac{Z Z'}{Z^2 + (2 k m/c) Z - a^2} = \frac{1}{r},
\ee
where $2 k m/c $ is the constant of integration.
The secondary integration yields
\be
(Z-Z_0)^{1-p} (Z+Z_1)^{1+p} = (C r)^2,
\ee 
where $C$ is the constant of integration, and
\ba
&&Z_0 = c^{-1} (\sqrt{(k m)^2 + c^2 a^2} - k m);\quad
Z_1 = c^{-1} (\sqrt{(k m)^2 + c^2 a^2} + k m), \\
&&p = k m [(k m)^2 + (c a)^2]^{-1/2} < 1. \nn
\ea
The equation (20) can be rewritten in the form:
\be
\left(
Z^2 + \frac{2 k m}{c} Z - a^2
\right)
\left(
\frac{Z+Z_1}{Z-Z_0}
\right)^p
= (C r)^2.
\ee

\begin{figure}
\centerline{
\epsfysize=.45\textwidth
\epsfbox{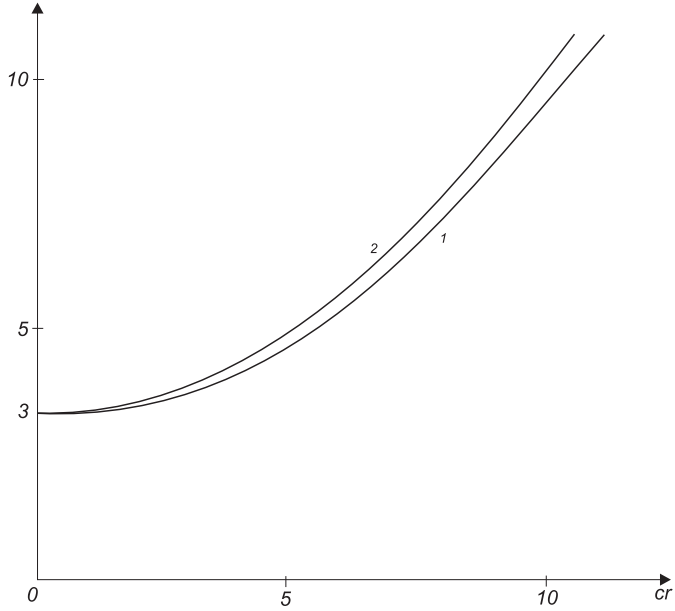}\\}
 \caption{{\sl 1} is $Z(r)$; {\sl 2} is a hyperbola}
\label{fig1}
\end{figure}

Equations (20) or (22) determine $Z$ as the function of $r$ depending
on the constants $C$, $m$ and parameter $a^2$.
At $Z\to \infty$ from (22) we obtain $Z\approx \pm C r$ so that $Z$ has
two branches.
Comparing this with (17) we see that it is necessary to suppose $C=c$
to choose the positive branch $Z$.
The curve $Z(r)$ is shown on Fig. \ref{fig1} at the following values
of parameters:
\be
c=1;~~k m= 0.158;~~ a=3.158;~~ p=0.05.
\ee
At $p\ll 1$ according to (22) the curve $Z(r)$ is close to the hyperbola:
\be
Z^2+ (2 k m/c) Z - a^2 = c^2 r^2,
\ee
which is shown on Fig. \ref{fig1} as well.

\begin{figure}
\centerline{
\epsfbox{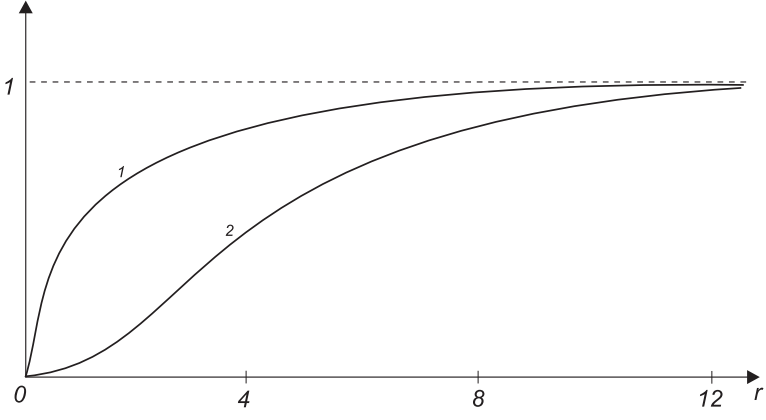}\\}
\caption{{\sl 1} is $e^\nu/c^2$; {\sl 2} is $e^\lambda$}
\label{fig2}
\end{figure}

By means of (15), (19) and (22) we obtain:
\ba
&& e^\nu = \frac{1}{r^2} 
\left(
      Z^2 + \frac{2 k m}{c} Z -a^2
\right) 
= c^2
\left(
      \frac{Z-Z_0}{Z+Z_1}
\right)^p, \nn\\
&&\\
&& e^\lambda = \frac{1}{Z^2} 
\left(
      Z^2 + \frac{2 k m}{c} Z -a^2
\right) 
= \frac{c^2 r^2}{Z^2}
\left(
      \frac{Z-Z_0}{Z+Z_1}
\right)^p. \nn
\ea
From here it is seen that for all $r\not= 0$ we have $e^\nu >0 $ and 
$e^\lambda > 0$.
In the point $r=0$ both functions tend to zero hence the gravitational 
radius is absent.
The curves (25) are represented on Fig. \ref{fig2} at the same values
of parameters (23).

As for the root of the determinant of metric tensor that
we find from (5) and (25):
\be
\sqrt{-g} = \frac{r \sin\vartheta}{Z}
\left(
      Z^2 + \frac{2 k m}{c} Z -a^2
\right) 
= \frac{c^2 r^3 \sin\vartheta}{Z}
\left(
      \frac{Z-Z_0}{Z+Z_1}
\right)^p. 
\ee

Approximately for $r\to 0$ we find from (20):
$Z-Z_0=\text{const}\cdot r^{2/(1-p)} + ...$, 
then from (25) we obtain
\be
e^\nu=\text{const}\cdot r^{2p/(1-p)} + ...;~~ 
e^\lambda=\text{const}\cdot r^{2/(1-p)} + ...
\ee
For $r\to \infty$ we find from (24)
\be
Z(r) = c r - (k m/c) + (k G^2 + k^2 m^2)/c^3 r  + ...
\ee
Then from (25)  up to the terms of order $1/r^2$ inclusive we obtain:
\be
e^\nu = c^2,~~ e^{-\lambda} =1 - 2 k m/c^2 r + 
(k G^2 + 2 k^2 m^2)/c^4 r^2  + ...
\ee

Thus, the metric has no Schwarzschild-like behavior in the sense that
it is non-symmetric with respect to $e^\nu$ and $e^{-\lambda}$.
At not very small scales the curving of spacetime reveals itself only 
in the spatial part of metric whereas time remains to be flat.
It is seen from Fig. \ref{fig2} as well.
Finally, from (29) we find the justification for the
designation $2 k m/c$ as an integration constant which was done in (19).

{\bf 3.} 
Let us consider now the expression for field $U$.
From (15), (19) and (25) we find:
\be
U'(r) =- \frac{c G}{r Z}  = -c G \frac{Z'}{Z^2 + (2 k m/c)Z -a^2}.
\ee
From here we obtain
\be
U(r) = \frac{c^2 G}{2\sqrt{k G^2 + k^2 m^2}}
\ln{\left(
          \frac{Z+Z_1}{Z-Z_0},
   \right)}
\ee
where the constant of integration is chosen in such a way 
that $U(\infty)=0$.
At $r\to 0$ we hence have:
\be
U=\text{const}\cdot\ln{(1/r)} + ...,
\ee
i.e., field has only the logarithmic singularity  at $r=0$ unlike
the ordinary result (3).

Further, from (6), (25), (26) and (30) we find for the tensor density
$\sqrt{-g}~ T^k_i$ the following expressions:
\be
T^0_0 \sqrt{-g} = (c^2 G^2/8 \pi) (\sin\vartheta/r Z),
\ee
\[
T^0_0 \sqrt{-g} = 
T^2_2 \sqrt{-g} = 
T^3_3 \sqrt{-g} = -
T^1_1 \sqrt{-g},
\]
which also differ completely from the ordinary result.
Hence for the total energy of static field we obtain:
\be
W = \frac{1}{c} \int T^0_0 \sqrt{-g}~ d x^1 d x^2 d x^3 =
(c G^2/ 2) \int d r/r Z (r).
\ee
The integral here is analogous to that in (30), (31), and if we take
$0$ and $\infty$ as limits that on the lower limit we obtain the 
divergence of order $W\sim \ln{(1/r)}$ at $r\to 0$, i.e., 
the total energy of the field of a point source diverges logarithmically.
It is interesting to point out that despite this circumstance the 
``classical charge radius'' remains to be the same as in the ordinary 
theory, $r_0 = G^2/ 2 m c$, as it can easily be made sure.
It is explained simply by the fact that the domain, where the curvature
of spacetime is sufficient, is much smaller than the 
``classical radius''.

Omitting elementary but bulk calculations we point out just that by
virtue of the substitution
\be
Z = c \rho - (k m/c) + (k G^2 + k^2 m^2)/ 4c^3 \rho,
\ee
the metric (5), (25) can be reduced to ``isotropic'' coordinates:
\[
d s^2 
= e^\nu d t^2 - e^\mu ( d \rho^2 + \rho^2 d \vartheta^2 + 
\rho^2 \Sin{2}{\vartheta}~ d \varphi^2) =
\]
\be
= e^\nu d t^2 - e^\mu ( d x^2 + d y^2 + d z^2),
\ee
where 
$\nu=\nu(\rho)$,
$\mu=\mu(\rho)$ and $\rho=\sqrt{x^2 + y^2 + z^2}$.
By means of the latter expression for $d s^2$ one can calculate,
also in an elementary but bulk way, the energy-momentum pseudo-tensor
of gravitational field $t^k_i$ and then the energy of total field
\be
P_0= c^{-1} \int (T^0_0 + t^0_0) \sqrt{-g} ~ d x\, d y\, d z.
\ee
As it is known \cite{ll} just $P_0$ rather than $W$ has physical sense.
We just point out that these calculations lead again to the
logarithmic divergence of total energy and to the same order of
``classical radius'' value as those above.

{\bf 4.}
In the case of scalar field with $\mass \not= 0$ the system (8) cannot
be resolved so simply as at $\mass = 0$.
However we are interested in the solution in the neighborhood
of the point $r=0$, and by direct substitution (27) and (32) into (8)
one can make sure that the solutions for the 
cases $\mass \not= 0$ and $\mass = 0$ coincide at $r\to 0$.
It is quite analogous to the result (2) of the ordinary theory.
The same result can be obtained without reference to the explicit form
of the solution for the case $\mass =0$.
Following from the invariance of system (8) with respect to the
transformation
\be
r\to \sigma r,\qquad
\mass\to \mass / \sigma,\qquad (\sigma = \text{const}),
\ee
it can be shown that the solution of this system has the form
(if then it is supposed  $\sigma=\mass$):
\be
U=U_0 V (r\mass),~~
e^\nu = e^{\nu_0} \varphi (r\mass),~~
e^\lambda = e^{\lambda_0} \psi (r\mass),
\ee
where $U_0$, $e^{\nu_0}$, $e^{\lambda_0}$ are the solutions for the
case $\mass =0$, and $V$, $\varphi$, $\psi$ are functions of a single
variable $(r\mass)$; these three functions $\to 1$ at $(r\mass)\to 0$.

Thus, we see that in the case of scalar mesostatic field 
(with $\mass \not= 0$ and $\mass = 0$) the considering of gravitational
effects at small distances from a source leads  to the results  
completely distinguished from the ordinary theory,
and these effects cannot be regarded as small.


\begin{thebibliography}{1}

\bibitem{ll}
L. Landau and E. Lifshitz,
\boo{Field Theory}{Litizdat}{Moskow}{1941}{}.


\end{thebibliography}
\end{document}